\documentclass[journal]{IEEEtran}
\usepackage{xcolor,soul,framed} 
\colorlet{shadecolor}{yellow}
\usepackage{graphicx}
\graphicspath{{../pdf/}{../jpeg/}}
\usepackage[cmex10]{amsmath}
\usepackage{array}
\usepackage{mdwmath}
\usepackage{mdwtab}
\usepackage{eqparbox}
\usepackage{url}
\usepackage{bm}
\usepackage{enumerate}
\usepackage{amsfonts}
\usepackage{nomencl}
\usepackage{etoolbox}
\usepackage{amssymb}
\usepackage{mathrsfs}
\usepackage{amsmath}
\usepackage{amsfonts,amssymb}
\usepackage{float}
\usepackage{stfloats}
\graphicspath{{figures/}}
\usepackage{calc}
\usepackage{tabularx}
\usepackage{fancybox}
\usepackage{subfig}
\usepackage{xcolor}
\usepackage[ruled,vlined]{algorithm2e}
\usepackage{bm}
\usepackage{bbm}

\usepackage{pdfpages}
\usepackage{svg}
\usepackage{booktabs}
\usepackage{tabu}
\usepackage{multirow}
\usepackage{makecell}
\usepackage{array}
\usepackage{hyperref}
\usepackage{amsthm}

\usepackage{dsfont}
\usepackage{bm}
\allowdisplaybreaks 
\usepackage[english]{babel}
\begin{document}
\captionsetup[figure]{name={Fig.},labelsep=period}
\bstctlcite{IEEEexample:BSTcontrol}
\title{Fast frequency response with heterogeneous communication delay management under the SCION Internet architecture}

\author{Jialun Zhang,~\IEEEmembership{Graduate Student Member,~IEEE,}
Felix Kottmann,
Jimmy Chih-Hsien Peng,~\IEEEmembership{Senior Member,~IEEE,}
Adrian Perrig,~\IEEEmembership{Fellow,~IEEE,}
    and Gabriela Hug,~\IEEEmembership{Senior Member,~IEEE}
        
\thanks{Jialun Zhang and Gabriela Hug are with Department of Information Technology and Electrical Engineering, ETH Zurich, 8092 Zurich, Switzerland.} 

\thanks{Felix Kottmann and Adrian Perrig are with the Department of Computer Science, ETH Zurich, 8092 Zurich, Switzerland.}%

\thanks{Jimmy Chih-Hsien Peng is with the Electrical and Computer Engineering, National University of Singapore, Singapore 117583, Singapore.}%
}

\maketitle

\begin{abstract}
System operators can increasingly exploit distributed energy resources (DERs) and controllable loads (CLs) to provide frequency response services. In conventional practice, communication between the system operator and flexible devices relies on the Border Gateway Protocol (BGP)-based Internet. However, existing BGP-based architectures face challenges in providing latency-guaranteed control, while direct private and proprietary communication networks lead to additional deployment and maintenance costs. In contrast, the SCION-based Internet architecture supports latency-minimum path selection, which makes it suitable for latency-sensitive frequency contingency services such as fast frequency response (FFR). Hence, this paper proposes a real-time reserve dispatch framework to optimally select a portfolio of flexible devices to deliver FFR services using the SCION-based Internet. First, an analytical expression of the system frequency dynamics with respect to heterogeneous communication latencies is derived. Next, a cyber-physical co-optimization model is formulated to jointly schedule communication paths and physical flexibility resources for real-time FFR provision. To improve the computation efficiency, we propose a heuristic FFR allocation algorithm to approximate the optimal response portfolio, integrating contributions from both DERs and CLs. Numerical case studies demonstrate the benefits of the proposed algorithm and its capability to approximate the optimality of the reserves allocation while significantly reducing the computation time. 
\end{abstract}
\IEEEpeerreviewmaketitle

\section{Introduction}
\IEEEPARstart{T}{he} transition towards a net-zero emission society requires the expansion of renewable energy sources and the phase-out of fossil fuel-based synchronous generators. Synchronous generations, however, play a vital role in keeping the system stable by supplying spinning reserves and preserving system inertia, features that inverter-based renewable energy sources inherently lack \cite{Daniel2020}. As the spinning reserves and system inertia decrease, power systems are more prone to experiencing significant and rapid frequency deviations during contingency events, which could result in power outages and even extensive system-wide blackouts, as demonstrated by the 2016 blackout incident in South Australia \cite{Austrablack2020}. 
 
In low inertia systems, FFR services can be implemented to rapidly respond and counteract frequency deviations caused by contingency events \cite{meng2019fast}. Compared to the slow reaction time of turbine-based generators, inverter-connected devices and load devices can deliver frequency support to the power grid within one second, making them highly suitable for FFR services \cite{Matamala2023}. Such flexible devices usually fall into two main categories: the first category includes storage-based DERs such as electric vehicles (EVs), battery energy storage systems (BESS), and uninterruptible power supplies (UPS). The second category consists of switchable CLs, which include household appliances such as refrigerators, heat pumps, and air conditioners \cite{talihati2024energy}. Nevertheless, coordinating these flexible devices poses notable difficulties due to the variation in their response dynamics, disparities in communication latencies, and the absence of a management entity overseeing activation requests from the system operators \cite{plaum2022aggregated}.


Specifically, system operators still face challenges in ensuring a consistent quality of reserve dispatch due to the heterogeneous activation delays of flexible devices \cite{hui2019modeling}. To address the latency uncertainties, Chen et al. integrates the modeling of communication delays into the power tracking controller to re-adjust power outputs from DERs in response to varying latencies in the event of contingencies \cite{chen2023scheduled}. In order to measure the economic impact of activation latency, Feng et al. calculates the range of operational costs by considering the highest and lowest latency observed in an aggregated frequency response model \cite{feng2022provision}. Wang et al. proposes a predictive model to replace the delayed or missing data to improve the convergence of distributed algorithms used by reserve dispatch entities \cite{wang2022asynchronous}. Furthermore, in order to prevent congestion within communication networks, Feng et al. suggests a scheduling strategy that selects a subset of DERs to receive updates from the distributed controller \cite{feng2023update}. These research efforts utilize the adaptability of flexible devices and controllers to counteract the unpredictability of latencies in the communication networks. 

\begin{figure}[!t]
    \centering
    \includegraphics[width=0.7\linewidth]{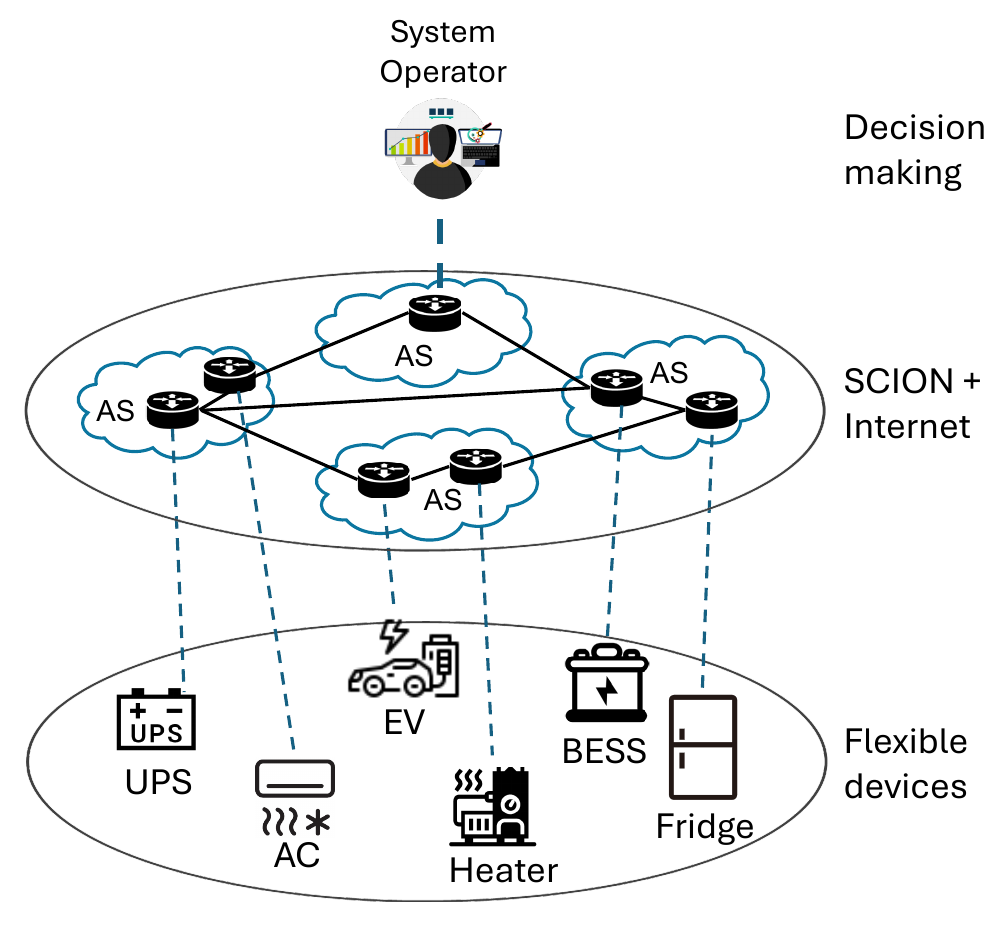}
    \caption{Structure of FFR dispatch architecture through the SCION Internet framework}
    \label{fig:overviewWP3}
\end{figure}

On the other hand, power systems require communication protocols that support scalability and interoperability to ensure seamless integration of a variety of flexible devices. Currently, the widely used protocols for power systems are based on TCP/IP and are incorporated into communication standards such as IEC 60870-5-104, IEC 61850, and Modbus that use proprietary communication networks \cite{kolenc2017performance}. However, costly proprietary communication networks are often not viable for widespread deployment of flexible devices. Hence, typically BGP-based public networks are employed. To address security concerns, Virtual Private Networks (VPN) are commonly employed. However, VPN-based communication methods face challenges in managing communication latencies because they do not prioritize low-latency routing \cite{da2017survey}. Extensive tests suggest that communication latencies using VPN connections may be further exacerbated by denial-of-service (DoS) attacks, primarily because the BGP protocol requires a significant period for re-convergence \cite{streun2022evaluating}. In contrast, SCION-enabled communication networks are capable of selecting the optimal set of communication paths measured by jitter, latencies and dropout rate and supports  communication over multiple paths \cite{labutkina2024multiobjective}. Moreover, by leveraging SCION's path-aware features, system operators can determine trusted routes resulting in minimum latency to interface with flexible devices \cite{10186417}. This method mitigates the effects of traffic congestion or path failures, guaranteeing the reliable delivery of FFR services. Figure~\ref{fig:overviewWP3} illustrates the proposed cyber-physical framework of the system operator to dispatch FFR services. Specifically, the system operator, deployed within an autonomous system (AS), utilizes a SCION-based Internet architecture to establish inter-domain connections with flexible devices located in other ASs.

Previous studies have concentrated on managing flexible devices or utilizing distributed control to mitigate the heterogeneous and time-varying activation latencies caused by communication networks. However, there remains a gap in the proactive management of activation latencies in the communication networks, which, alongside reserve capacity, influences frequency deviation. Consequently, this paper proposes a cyber-physical co-optimization approach for the system operator by employing the SCION internet to promptly and reliably deliver FFR services. The proposed approach specifically utilizes the multi-path routing features of SCION to identify optimal routes that reduce communication delays between the system operator and flexible resources. A decrease in communication delays allows the system operator's dispatch command to rapidly reach the flexible devices, thereby delivering high-quality FFR services. This research is the first to integrate the SCION Internet architecture into energy management strategies and to evaluate the potential for SCION-enabled reserve dispatch operations. It offers the means for future cyber-physical designs of FFR dispatch using public Internet connections. 

The major contributions of this paper are as follows:
\begin{enumerate}
    \item We propose a cyber-physical operational architecture for the system operator to dispatch FFR services with SCION. This operational architecture enables the system operator to dynamically re-route its communication path in response to real-time communication latencies which in turn ensures swift and reliable reserve provision from DERs and CLs in case of contingency events.
    \item An analytical frequency nadir expression is derived taking into account both a high order aggregated multi-machine system frequency response model and the dynamics of DERs and CLs under heterogeneous communication latencies. Specifically, the complex components from the frequency response model are transformed into real components in the analytical nadir expression to simplify the integration into an optimization problem. 
    \item We propose a heuristic method for routing and reserve allocation to tackle the computational complexity inherent in the nonlinear cyber-physical reserve dispatch problem formulated. Specifically, the dispatch problem is solved iteratively utilizing a heuristic algorithm for reserve allocation, which is based on the order of communication latencies. Moreover, considering the steady-state frequency condition, a set of computation acceleration methods are employed to enhance the convergence of the heuristic algorithm.
\end{enumerate}

The rest of the paper is structured as follows. Sec~\ref{Investment} presents the derivations of the frequency dynamic models based on the modeling of synchronous generators, DERs and controllable loads. In Sec~\ref{commresourc}, we formulate the complete cyber-physical FFR dispatch problem. Then, Sec~\ref{sec:heurisalloc} presents the heuristic reserve allocation algorithm to iteratively choose the minimum required reserve in the dispatch portfolio. In Sec~\ref{Casestudies}, we present case studies to verify the effectiveness of the proposed framework, followed by the findings and conclusions in Sec~\ref{conclusion}.

\section{Frequency dynamic modelling}\label{Investment}

\begin{figure}[!t]
    \centering
    \includegraphics[width=1\linewidth]{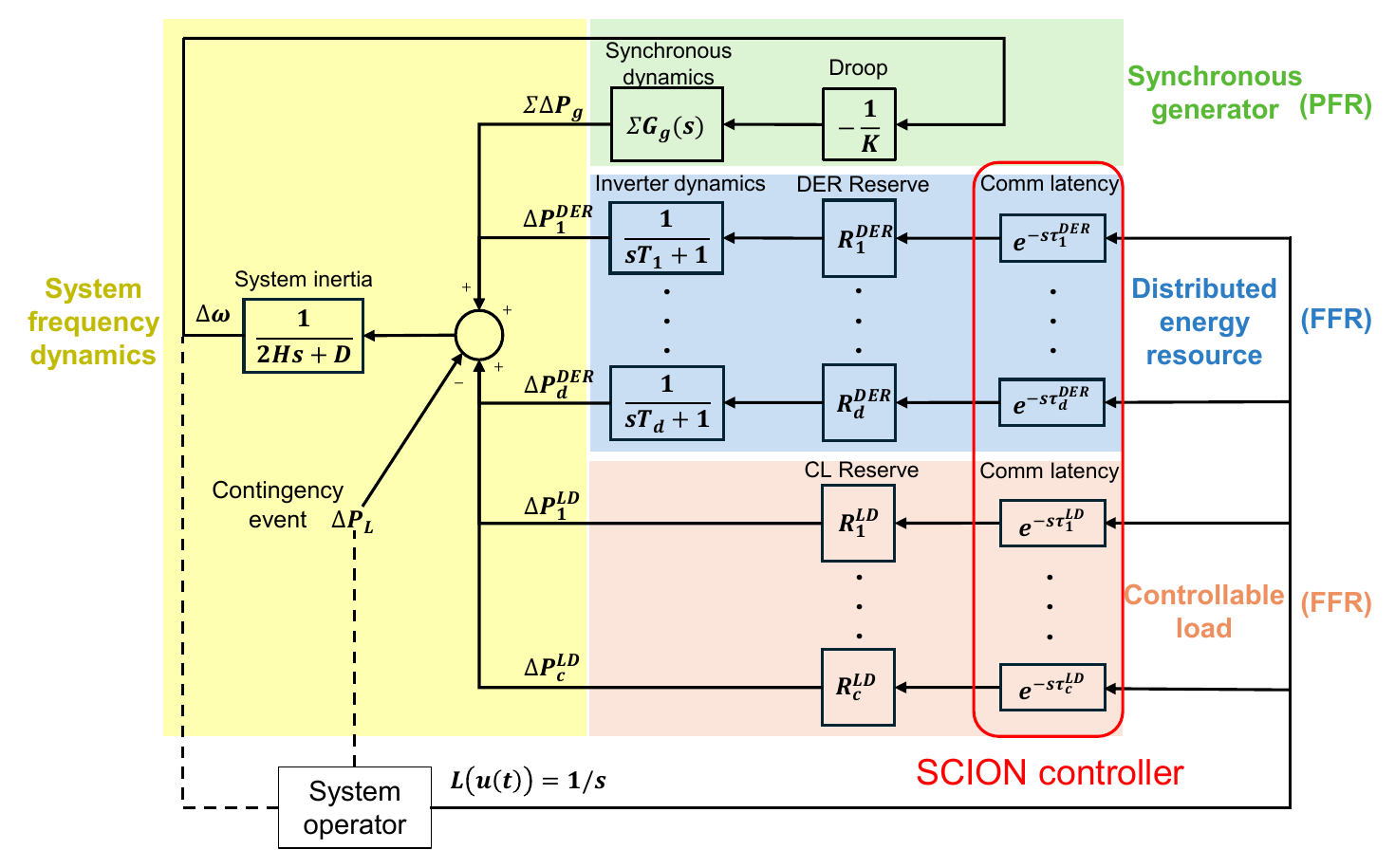}
    \caption{Fast frequency response with diverse communication latencies in a cyber-physical system architecture using SCION internet}
    \label{fig:Frequencydev}
\end{figure}

The block diagram of the system frequency dynamics model is shown in Fig.~\ref{fig:Frequencydev}. In this model, we define three types of response elements: synchronous generator $g\in \mathcal{G}$, inverter-based DER $d\in \mathcal{D}$ and controllable load (CL) $c\in \mathcal{C}$. It is assumed that synchronous generators are used to provide primary frequency response (PFR), whereas DERs and CLs are used to provide FFR service in response to a power contingency event $\Delta P_L$.

For synchronous generators, the PFR is typically implemented by droop control with parameter $K$ which adjusts the PFR reserve output $\sum \Delta P_g$ in response to the power mismatch. Here, $\sum G_g(s)$ represents the generic expression of the aggregated dynamic response of the synchronous generators. 

For inverter-based DERs and controllable loads, the fast frequency response (FFR) service is activated through the activation command $L(\delta)$ (i.e., impulse signal in frequency domain) from the system operator. However, due to the propagation via internet-based communication, the activation command experiences a time delay of $\tau_d^{DER}$ for any DER $d\in \mathcal{D}$ and a time delay of $\tau_c^{LD}$ for any controllable load $c\in \mathcal{C}$, which are actively monitored and managed by the SCION controller. After receiving the activation command, all inverter-based DERs $d\in \mathcal{D}$ and controllable loads $c\in \mathcal{C}$ react and, if needed, could ramp up their power to their maximum reserve capacity $R_d^{DER}$ and $R_c^{LD}$, respectively. For inverter-based DERs, the inverter dynamics blocks characterize their response dynamics with the respective response time $T_d$. 

The general frequency response of the power system is based on the swing equation characterized by the inertia of the system, i.e., $H=\sum_{g}^{G}S_g\cdot H_g/S_{sym}$, and the damping coefficient $D$. The system operator constantly monitors the frequency deviation $\Delta \omega$ and the generation-demand mismatch caused by a contingency event $\Delta P_L$. The modeling of the specialized communication network between the system operator and the monitoring devices is not the scope of this work and, therefore, omitted. In this context, the system operator is broadly defined as the authority responsible for determining which flexible devices should be activated, irrespective of whether they are transmission or distribution system operators.

\section{Frequency response analytical form with multi-delays}
By aggregating the external power injection $\Delta P = \sum_{d\in \mathcal{D}} \Delta P_d^{DER}+\sum_{c\in \mathcal{C}} \Delta P_c^{LD}-\Delta P_L$, the transfer function of frequency deviations $\Delta \omega$ can be written as follows:
\begin{equation}
    \frac{\Delta \omega}{\Delta P(s)} = \frac{(2Hs+D)^{-1}}{1+\sum G_g(s)K^{-1}(2Hs+D)^{-1}}\label{eq:transferori}
\end{equation}

Typically, the system is a causal system and its transfer function is rational with distinct poles. In this case, the transfer function~\eqref{eq:transferori} can be defactorized into a sum of first order systems \cite{feng2022provision}:
\begin{equation}\label{eq:WP3normalpoles}
    \frac{\Delta \omega}{\Delta P(s)}=\sum_{(\alpha_i, p_i)\in \mathcal{P}} \frac{\alpha_i}{s+p_i}
\end{equation}
where in the set of poles and residues $\mathcal{P}$, each pole $p_i$ and residue $\alpha_i$ can be a real or complex conjugate pair in the frequency domain. However, since some poles and residues may involve complex numbers (i.e., $\alpha_u \in \mathbb{C}, p_u \in \mathbb{C}$ where $(\alpha_u, p_u)\in \mathcal{P}$) that are hard to optimize in the time domain, we reformulate the expression~\eqref{eq:WP3normalpoles} as follows:
\begin{equation}\label{eq:WP3normalpolesnew}
    \frac{\Delta \omega}{\Delta P(s)}= \sum_{\substack{(\alpha_u, p_u)\in \mathcal{P} \\ \alpha_u \in \mathbb{C} \\ p_u \in \mathbb{C}}} \frac{\Re(\alpha_u)+j\Im(\alpha_u)}{s+\Re(p_u)+j\Im(p_u)}+\sum_{\substack{(\alpha_r, p_r)\in \mathcal{P} \\ \alpha_r \in \mathbb{R} \\ p_r \in \mathbb{R}}} \frac{\alpha_r}{s+p_r}
\end{equation}
where $\alpha_r$ and $p_r$ are real residues and poles whereas $\alpha_u$ and $p_u$ include complex conjugate pairs of the residues and poles, i.e., $a_i/(s+p_i)+a_i^*/(s+p_i^*)$ . $\Re(\cdot)$ and $\Im(\cdot)$ denotes the real and imaginary component of the complex term, respectively. All poles and residues form the set $\mathcal{P}$. As a result, the transfer function can be transformed into a sum of second order systems and first order systems:
\begin{align}
\begin{split}
    \frac{\Delta \omega}{\Delta P(s)} = \sum_{\substack{(\alpha_u, p_u)\in \mathcal{P} \\ \alpha_u \in \mathbb{C} \\ p_u \in \mathbb{C}}} 
    \Bigg[
    \frac{2\Re(\alpha_u)(s + \Re(p_u))}{(s + \Re(p_u))^2 + \Im(p_u)^2} 
    \\ + \frac{2\Im(\alpha_u)\Im(p_u)}{(s+\Re(p_u))^2 + \Im(p_u)^2}
    \Bigg] + \sum_{\substack{(\alpha_r, p_r)\in \mathcal{P} \\ \alpha_r \in \mathbb{R} \\ p_r \in \mathbb{R}}} \frac{\alpha_r}{s+p_r}
\end{split} \label{eq:residuecomplex}
\end{align}

Thus, the total system frequency deviation $\Delta \omega(s)$ can be determined by the combined contribution from DERs, CLs and the power mismatch due to contingency event $\Delta P_L$:
\begin{equation}
    \Delta \omega(s) =  \sum_{d\in \mathcal{D}} \Delta \omega_d^{DER} (s)+\sum_{c\in \mathcal{C}} \Delta \omega_c^{LD} (s)-\Delta\omega_L(s)\label{eq:omegasum}
\end{equation}
where $\Delta \omega_d^{DER} (s)$ denotes the frequency response from DER $d$, $\Delta \omega_c^{LD} (s)$ represents the frequency response from CL $c$ and $\Delta\omega_L(s)$ is the frequency deviation caused by the contingency event $\Delta P_L$.

Meanwhile, the FFR power injection once the system operator issues activation signal $L(u(t))=1/s$ can be expressed as:
\begin{align}
\begin{split}
    &\Delta P_d^{DER} = \frac{R_d^{DER}e^{-s \tau_d^{DER}}}{s(sT_d+1)} \\
      &\Delta P_c^{LD} =  \frac{R_c^{LD}e^{-s \tau_c^{LD}}}{s}
\end{split}\label{eq:pexpress}
\end{align}

Based on~\eqref{eq:transferori} and~\eqref{eq:pexpress}, the frequency deviations attributed to DERs, CLs and power mismatch in the frequency domain are as follows:
\begin{align}
\begin{split}
    \sum_{d\in \mathcal{D}} \Delta \omega_d^{DER}(s) &= \sum_{d\in \mathcal{D}} \frac{R_d^{DER}e^{-s \tau_d^{DER}}}{s(sT_d+1)} \frac{\Delta \omega}{\Delta P(s)}\\
      \sum_{c\in \mathcal{C}}\Delta \omega_c^{LD}(s) &=  \sum_{c\in \mathcal{C}} \frac{R_c^{LD}e^{-s \tau_c^{LD}}}{s}  \frac{\Delta \omega}{\Delta P(s)}\\
       \Delta\omega_L(s) &= \frac{\Delta P_L}{s} \frac{\Delta \omega}{\Delta P(s)}
        \end{split}\label{eq:frequeninsep}
\end{align}

By combining~\eqref{eq:omegasum} and~\eqref{eq:frequeninsep}, the frequency response in the time domain can be derived as:
\begin{equation}\label{eq:wexpress}
    \Delta \omega(t) =  \sum_{d\in \mathcal{D}} \Delta \omega_d^{DER} (t)+\sum_{c\in \mathcal{C}} \Delta \omega_c^{LD} (t)-\Delta\omega_L(t)
\end{equation}
where each component is given in~\eqref{eq:finalfreqdynamic}. Here we briefly explain the structure of~\eqref{eq:finalfreqdynamic} whereas more detailed derivations are given in the Appendix: to enable a simpler inverse Laplace transformation, the non-zero pole of $\Delta P_d^{DER}$ in \eqref{eq:pexpress}, i.e., $1/(sT_d+1)$, is also included as a pole in the residue decomposition, so that the time shift ($e^{-s \tau}$) and the integral component ($1/s$) share the same structure for the frequency response for DERs and CLs. The set of poles and residues for the expression of DERs $\mathcal{P}_{DER}$ therefore contains an additional pole at $-\frac{1}{T_d}$. Therefore, the term $\mathcal{S}$ in~\eqref{eq:finalfreqdynamic} for expressions $f_1$ and $f_2$ denotes a symbolic expression of residues and poles, which is substituted by $\mathcal{P}_{DER}$ for the frequency response of DER $d$ (i.e., $\Delta \omega_d^{DER}(t)$) and is substituted by the original set of poles and zeros $\mathcal{P}$ from~\eqref{eq:WP3normalpoles} for the frequency response of CL $c$ (i.e., $\Delta \omega_c^{LD}(t)$) as well as for the frequency response of the power mismatch $\Delta \omega_L(t)$. 

\begin{figure*}[!b]
\hrulefill
\begin{align}
    \begin{split}
        \sum_{d\in \mathcal{D}} \Delta \omega_d^{DER}(t) &=  \sum_{d\in \mathcal{D}}R_d^{DER} \Bigg[
        \sum_{\substack{(\alpha_r, p_r)\in \mathcal{P}_{DER} \\ \alpha_r \in \mathbb{R} \\ p_r \in \mathbb{R}}} 
        \frac{\alpha_r}{p_r} \left(1-e^{-p_r(t-\tau_d^{DER})} \right) \\
        & \quad + f_1(t - \tau_d^{DER}; \mathcal{P}_{DER}) 
        + f_2(t - \tau_d^{DER}; \mathcal{P}_{DER}) \Bigg] u(t-\tau_d^{DER}) \\
        \sum_{c\in \mathcal{C}}\Delta \omega_c^{LD}(t) &=  \sum_{c\in \mathcal{C}}R_c^{LD} \Bigg[
        \sum_{\substack{(\alpha_r, p_r)\in \mathcal{P} \\ \alpha_r \in \mathbb{R} \\ p_r \in \mathbb{R}}} 
        \frac{\alpha_r}{p_r} \left(1-e^{-p_r(t-\tau_c^{LD})} \right)  + f_1(t - \tau_c^{LD}; \mathcal{P}) 
        + f_2(t - \tau_c^{LD}; \mathcal{P}) \Bigg] u(t-\tau_c^{LD}) \\
        \Delta \omega_L(t) &= \Delta P_L \Bigg[
        \sum_{\substack{(\alpha_r, p_r)\in \mathcal{P} \\ \alpha_r \in \mathbb{R} \\ p_r \in \mathbb{R}}} 
        \frac{\alpha_r}{p_r} \left(1-e^{-p_r t}\right) + f_1(t; \mathcal{P}) 
        + f_2(t; \mathcal{P}) \Bigg] u(t) \\
        f_1(t; \mathcal{S}) &= \sum_{\substack{(\alpha_u, p_u)\in \mathcal{S} \\ \alpha_u \in \mathbb{C} \\ p_u \in \mathbb{C}}} 
        \frac{2\Re(\alpha_u)}{\Re(p_u)^2 + \Im(p_u)^2} 
        \Big[ -e^{-\Re(p_u)t} \Re(p_u) \cos(\Im(p_u)t)  + \Re(p_u) + e^{-\Re(p_u)t} \Im(p_u) \sin(\Im(p_u)t) \Big] \\
        f_2(t; \mathcal{S}) &= \sum_{\substack{(\alpha_u, p_u)\in \mathcal{S} \\ \alpha_u \in \mathbb{C} \\ p_u \in \mathbb{C}}} 
        \frac{2\Im(\alpha_u)}{\Re(p_u)^2 + \Im(p_u)^2} 
        \Big[ -e^{-\Re(p_u)t} \Re(p_u) \sin(\Im(p_u)t)  + \Im(p_u) - e^{-\Re(p_u)t} \Im(p_u) \cos(\Im(p_u)t) \Big]
    \end{split} \label{eq:finalfreqdynamic}
\end{align}
\end{figure*}

Generally, with the RoCoF at the nadir point being equal to zero, we have the following formulation for $t_{nad}$:
\begin{align}\label{eq:tnadform}
\begin{split}
    \Delta \dot{\omega}(t_{nad}) =&  \sum_{d\in \mathcal{D}} \Delta \dot{\omega}_d^{DER} (t_{nad})+\sum_{c\in \mathcal{C}} \Delta \dot{\omega}_c^{LD} (t_{nad})\\
    &-\Delta\dot{\omega}_L (t_{nad}) = 0
\end{split}
\end{align}
for which the derivative components are given in~\eqref{eq:finalfreqdynamicderiv}. Once solved, we can substitute $t_{nad}$ back into \eqref{eq:wexpress} to obtain the frequency nadir output $\Delta w(t_{nad})$, namely
\begin{figure*}[!t]
\begin{align}
\begin{split}
    \sum_{d\in \mathcal{D}} \Delta \dot{\omega}_d^{DER}(t) &= \sum_{d\in \mathcal{D}}R_d^{DER}\Bigg[\sum_{\substack{(\alpha_r, p_r)\in \mathcal{P}_{DER} \\ \alpha_r \in \mathbb{R} \\ p_r \in \mathbb{R}}} \alpha_r e^{-p_r(t-\tau_d^{DER})} + \dot{f}(t-\tau_d^{DER}; \mathcal{P}_{DER}) \Bigg]u(t-\tau_d^{DER})\\
    \sum_{c\in \mathcal{C}}\Delta \dot{\omega}_c^{LD}(t) &=   \sum_{c\in \mathcal{C}}R_c^{LD}\Bigg[\sum_{\substack{(\alpha_r, p_r)\in \mathcal{P} \\ \alpha_r \in \mathbb{R} \\ p_r \in \mathbb{R}}} \alpha_r e^{-p_r(t-\tau_c^{LD})} + \dot{f}(t-\tau_c^{LD}; \mathcal{P}) \Bigg]u(t-\tau_c^{LD})\\
    \Delta\dot{\omega}_L(t) &= \Delta P_L\Bigg[\sum_{\substack{(\alpha_r, p_r)\in \mathcal{P} \\ \alpha_r \in \mathbb{R} \\ p_r \in \mathbb{R}}} \alpha_r e^{-p_r t} + \dot{f}(t; \mathcal{P})\Bigg]u(t) \\
    \dot{f}(t; \mathcal{S}) &= \sum_{\substack{(\alpha_u, p_u)\in \mathcal{S} \\ \alpha_u \in \mathbb{C} \\ p_u \in \mathbb{C}}} 2 \cdot e^{-\Re(p_u)t}\Big[ \Im(\alpha_u)\sin(\Im(p_u)t)  + \Re(\alpha_u)\cos(\Im(p_u)t) \Big]
\end{split}\label{eq:finalfreqdynamicderiv}
\end{align}
\hrulefill
\end{figure*}

\begin{equation}\label{eq:wexpressnadir}
\begin{split}
        \Delta \omega(t_{nad}) = & \sum_{d\in \mathcal{D}} \Delta \omega_d^{DER} (t_{nad})+\sum_{c\in \mathcal{C}} \Delta \omega_c^{LD} (t_{nad})\\&-\Delta\omega_L(t_{nad})
\end{split}
\end{equation}


\section{Cyber-physical co-optimization model for FFR provision}\label{commresourc}
The system operator seeks to minimize the remuneration expenses for activating FFR reserves while simultaneously minimizing the overall communication latency by opting for routes with the lowest latency thereby ensuring responsive FFR services for balancing a power contingency event. Let $\mathcal{A}_d$ and $\mathcal{A}_c$ be the set of available communication paths for each DER $d$ and controllable load $c$, respectively. Therefore, the following cyber-physical optimization problem is formulated: 
\begin{subequations}\label{WP3VPPoperation}
\begin{align}
     \min_{\substack{R_d^{DER},R_c^{LD},\\ z_d^{DER}, z_c^{LD}, t_{nad}}}  & \begin{aligned}
         C_{rr}\left(\sum_{d\in \mathcal{D}}{z_d^{DER}R_d^{DER}}+\sum_{c\in \mathcal{C}}{z_c^{LD}R_c^{LD}}\right)\\ +\sum_{d\in \mathcal{D},c\in \mathcal{C}}(\tau_d^{DER}+\tau_c^{LD}) \end{aligned}\\
 \text{s.t.} \quad & \Delta \omega (t_{nad}) \leq \Delta \omega^{max}, \quad  \Delta \dot{\omega}(t_{nad})=0 \label{con:complexnadir}\\
 & \sum_{d\in \mathcal{D}}{R_d^{DER}} +\sum_{c\in \mathcal{C}}{R_c^{LD}} \geq \Delta P_L \label{con:steadystatecon}\\
& R_d^{DER} \leq z_d^{DER} \cdot R_{d}^{DERmax} \quad : \forall d\in \mathcal{D} \label{con:WP3RDupp}\\
& R_c^{LD} \leq z_c^{LD} \cdot R_{c}^{LDmax} \quad : \forall{c} \in \mathcal{C} \label{con:WP3Rcupp}\\
    &\sum_{pa \in \mathcal{A}_d} \eta_d^{pa} = 1 \quad : \forall d\in \mathcal{D} \label{con:pathdone} \\
    &\sum_{pa \in \mathcal{A}_c} \eta_c^{pa} = 1 \quad : \forall{c} \in \mathcal{C} \label{con:pathcone} \\
    &\sum_{pa \in \mathcal{A}_d} \tau_d^{pa} \eta_d^{pa} \leq \tau_d^{DER} \quad : \forall d\in \mathcal{D} \label{WP3con:pathselder}\\
    &\sum_{pa \in \mathcal{A}_c} \tau_c^{pa} \eta_c^{pa} \leq \tau_c^{LD} \quad : \forall{c} \in \mathcal{C} \label{WP3con:pathselCL}\\
    & \eta_d^{pa} \in \{0,1\} \quad : \forall{pa} \in \mathcal{A}_d, \forall d\in \mathcal{D}  \label{WP3con:integerder}\\
    & \eta_c^{pa} \in \{0,1\} \quad : \forall{pa} \in \mathcal{A}_c, \forall c \in \mathcal{C}\label{WP3con:integerc}\\
    & z_d^{DER} \in \{0,1\} \quad : \forall d\in \mathcal{D} \label{con:binaryder}\\
    & z_c^{LD} \in \{0,1\} \quad : \forall{c} \in \mathcal{C} \label{con:binarycl}
\end{align}
\end{subequations} 
where $C_{rr}$ denotes the remuneration rate paid by the system operator to DERs and CLs for the activation of FFR services, $z_d^{DER}$ and $z_c^{LD}$ denote the activation decision variables for DER $d$ and CL $c$, $\tau_d^{DER}$ and $\tau_c^{LD}$ represent the communication latencies between the system operator and DER $d$ and CL $c$, respectively. Constraint~\eqref{con:complexnadir} enforces that the frequency nadir needs to be within the maximum frequency deviation limit $\Delta \omega^{max}$. Constraint~\eqref{con:steadystatecon} ensures sufficient steady-state balancing reserves for a contingency event $\Delta P_L$. Constraints~\eqref{con:WP3RDupp} to~\eqref{con:WP3Rcupp} restrict the activated reserves to their upper bounds for $R_{d}^{DERmax}$ and $R_{c}^{LDmax}$ for DER $d$ and CL $c$, respectively, considering both the maximum output capacity of the devices and feed-in limits for the power grid, respectively. Combined with the respective terms in the objective function, constraints \eqref{con:pathdone} to \eqref{WP3con:integerc} ensure the selection of the communication paths with the lowest latency between the system operator and flexible devices. Specifically, constraints~\eqref{con:pathdone} to~\eqref{con:pathcone} ensure that in the set of available paths $\mathcal{A}_d$ for DERs and the set of available paths $\mathcal{A}_c$ for CLs, only one path is selected for DER $d$ and CL $c$. Finally, constraints \eqref{WP3con:integerder} to \eqref{WP3con:integerc} define the integer decision variables for the path selection process, while constraints~\eqref{con:binaryder} to~\eqref{con:binarycl} define the decision variables for the activation decision of DER $d$ and CL $c$.

Since the frequency nadir constraint in~\eqref{con:complexnadir} is nonlinear, the cyber-physical optimization problem~\eqref{WP3VPPoperation} is a mixed-integer nonlinear programming (MINLP) problem. This problem is computationally intensive and challenging to solve due to its NP-hard nature.

\section{Lowest-latency routing and heuristic reserve allocation algorithm}\label{sec:heurisalloc}
In this section, we introduce a heuristic method to decompose the solution process of the cyber-physical co-optimization model~\eqref{WP3VPPoperation} into sequential steps. In Section~\ref{WP3IVA}, we employ the deadline-aware multipath transport protocol (DMTP) developed for SCION to predict delay times and choose the path with the lowest latency for the communication between the system operator and its flexible devices. In Section \ref{WP3IVB}, an equivalent step response model is introduced to represent the frequency response of DERs. In Section \ref{WP3IVC}, the activation latencies for both DERs and CLs are sorted in ascending order to establish the sequence for iterative reserve allocation. Finally, Section~\ref{WP3IVD} describes the reserve allocation procedure for each iteration, while Section~\ref{WP3IVE} presents the computation acceleration techniques to reduce the number of iterations. 

\subsection{Lowest-latency routing in SCION communication network}\label{WP3IVA}
The first step is to optimize the communication routes from the system operator control center to the DERs for minimal latency. The routing problem can be formulated in accordance with problem~\eqref{WP3VPPoperation} as:
\begin{align} 
    & \min \sum_{d\in \mathcal{D},c\in \mathcal{C}}(\tau_d^{DER}+\tau_c^{LD})\\
     & \text{s.t.} \quad \eqref{con:pathdone} - \eqref{WP3con:integerc} \nonumber
\end{align}

For the SCION Internet architecture, the lowest-latency routing is realized by the DMTP developed in \cite{10186417}. Specifically, the system operator first identifies all disjoint paths between the AS hosting the system operator and the ASs hosting DERs (i.e., $\mathcal{A}_d$) and CLs (i.e., $\mathcal{A}_c$). Then, the active probing scheme measures the latency of each path and selects the one with the lowest communication latency between the system operator and any DER $d$ (i.e., $\tau_d^{DER}$) and CL $c$ (i.e., $\tau_c^{LD}$). In case of latency changes, DMTP will leverage its built-in failover mechanism to promptly reroute traffic to the next lowest-latency path. This method of prompt re-routing guarantees that communications between the system operator and the flexible devices occurs always with the lowest possible latency.

\subsection{Equivalent step response model of DER}\label{WP3IVB} 
\begin{figure}[!t]
    \centering
    \includegraphics[width=0.75\linewidth]{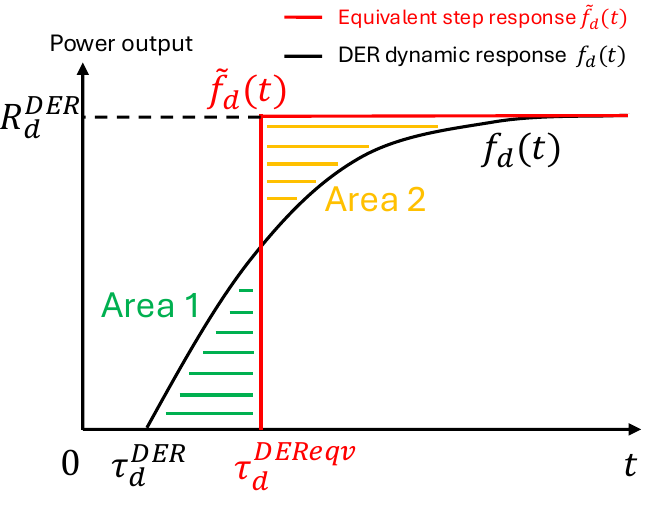}
    \caption{Equivalent step response model for frequency dynamics of DER $d$}
    \label{fig:equivalentderWP3}
\end{figure}
The second step is to derive the equivalent step response model for DERs. The first order dynamic response from DER $d$ in the time domain can be expressed as:
\begin{equation}
    f_{d}(t)=R_d^{DER}\left(1-e^{-(t-\tau_d^{DER})/T_d}\right)u(t-\tau_d^{DER})
\end{equation}

On this basis, an equivalent step response function for DER $d$ is introduced as follows: 
\begin{equation}
    \Tilde{f}_{d}(t)=R_d^{DER}u(t-\tau_{d}^{DEReqv})
\end{equation}
where $\tau_{d}^{DEReqv}$ denotes the equivalent latency of the step response model for DER $d$. To compute the equivalent time latency $\tau_{d}^{DEReqv}$, we apply the law of conservation of energy, requiring that the energy output from the DER dynamics model matches that of the step response model. As illustrated in Fig.~\ref{fig:equivalentderWP3}, this is achieved by requiring the equivalence of Area 1 and Area 2, implying that the energy produced by the dynamic response model prior to the activation of the equivalent step response model must be balanced by the surplus of energy produced by the equivalent step response model after activation. As a result, the following equation needs to be satisfied:
\begin{equation}\label{WP3:eqvmodel}
\begin{split}
    &R_d^{DER}\int_{\tau_d^{DER}}^{\tau_{d}^{DEReqv}}\left(1-e^{-(t-\tau_d^{DER})/T_d}\right)u(t-\tau_d^{DER})dt =\\
    &\int_{\tau_{d}^{DEReqv}}^{\infty}R_d^{DER}u(t-\tau_{d}^{DEReqv})dt -\\
    &\int_{\tau_{d}^{DEReqv}}^{\infty}R_d^{DER}(1-e^{-(t-\tau_d^{DER})/T_d})u(t-\tau_d^{DER})dt
\end{split}
\end{equation}

Solving~\eqref{WP3:eqvmodel} yields the equivalent time latency of the step response model of DER $d$ as: 
\begin{equation}
    \tau_d^{DEReqv}=\tau_d^{DER}+T_d
\end{equation}

\subsection{Ascending latency sorting function}\label{WP3IVC} 
Based on the derivation of the equivalent time latency of DERs, we define the set of latencies from DERs as $\mathcal{T}^{DEReqv} = \{\tau_d^{DEReqv} \mid d \in \mathcal{D}\}$ and set of latencies for CLs as $\mathcal{T}^{LD} = \{\tau_c^{LD} \mid c \in \mathcal{C}\}$. Then, a latency set $\mathcal{T}^{all}$ concatenating the elements from $\mathcal{T}^{DEReqv}$ and $\mathcal{T}^{LD}$ while maintaining the sequence is introduced:
\begin{align}
    \mathcal{T}^{\text{all}} = \{ (\tau_{id}, id) \mid id = 1, 2, \dots, |\mathcal{D}| + |\mathcal{C}| \}
\end{align}
where ${id}$ represents the index of latencies in the concatenated latency set. In this concatenated latency set, $\tau_{id}$ denotes a DER when $1 \leq id \leq |\mathcal{D}|$ and corresponds to a CL when $|\mathcal{D}|+1 \leq id \leq |\mathcal{D}|+|\mathcal{C}|$.

We then introduce a sorting function $\sigma_{so}$ to rearrange the latencies from the concatenated latency set $\mathcal{T}^{all}$ in an ascending order:
\begin{align}
\begin{split}
      \sigma_{so}(\mathcal{T}^{all})=&\{ (\tau_{(1)}, id_{(1)}), (\tau_{(2)}, id_{(2)}), \\&\dots, (\tau_{(|\mathcal{D}| + |\mathcal{C}|)}, id_{(|\mathcal{D}| + |\mathcal{C}|)}) \}  
\end{split}
\end{align}
where $\tau_{(1)} \leq \tau_{(2)} \leq \dots \leq \tau_{(|\mathcal{D}| + |\mathcal{C}|)}$. In other words, $\tau_{(k)}$ represents the $k$-th lowest latency element in the concatenated latency set $\mathcal{T}^{all}$ and $id_{(k)}$ denotes the index of the $k$-th lowest latency element in the unsorted set.

\subsection{Iterative reserve allocation}\label{WP3IVD}
This section illustrates how to iteratively allocate the reserve from DERs and CLs according to the ascending order of latencies. First, we initialize an empty set for the DERs to be activated as $\mathcal{D}^{act}$ and an empty set for CLs to be activated as $\mathcal{C}^{act}$. Then, we start the iteration from the first element of the concatenated latency set. In any iteration $k$, we determine whether and which DER or CL is activated based on the following two conditions: first, if $1 \leq id_{(k)} \leq |\mathcal{D}|$, activate DER $d=id_{(k)}$ with its maximum reserve capacity $R_{d}^{DERmax}$ and incorporate its index in the activation latency set for DERs i.e., 
\begin{align}
     \mathcal{D}^{act}_{(k)} \cup id_{(k)}
\end{align}

The second condition is if $|\mathcal{D}| < id_{(k)} \leq |\mathcal{D}|+|\mathcal{C}|$, CL at index $c=id_{(k)}-|\mathcal{D}|$ is activated with its maximum output $R_{c}^{LDmax}$ and its communication latency is incorporated in the CL activation latency set:
\begin{align}
    \mathcal{C}^{act}_{(k)} \cup \left(id_{(k)}-|\mathcal{D}|\right)
\end{align}

In each iteration, we obtain a set of reserves to be activated by DERs as $\mathcal{R}^{DERact}_{(k)}=\{R_{da}^{DERmax}\mid da \in \mathcal{D}^{act}\}$ with its latency set $\mathcal{T}_{(k)}^{DERact}=\{\tau_{da}^{DER}\mid da \in \mathcal{D}^{act}\}$, and the set of reserves to be activated by CLs $\mathcal{R}^{LDact}_{(k)}=\{R_{ca}^{LDmax}\mid ca \in \mathcal{C}^{act}\}$ with its latency set as $\mathcal{T}^{LDact}_{(k)}=\{\tau_{ca}^{LD}\mid ca \in \mathcal{C}^{act}\}$. We then calculate the nadir time numerically by substituting these reserves and latencies to be activated into~\eqref{eq:finalfreqdynamicderiv} and solving:
\begin{equation}\label{con:sumtadcalculate}
\begin{split}
    &\Delta \dot{\omega}_{sum}^{DER} (\mathcal{R}^{DERact}_{(k)},\mathcal{T}^{DERact}_{(k)}, t_{nad}) +\\ &\Delta \dot{\omega}_{sum}^{LD} (\mathcal{R}^{LDact}_{(k)},\mathcal{T}^{LDact}_{(k)}, t_{nad})-
    \dot{\omega}_L (t_{nad}) = 0
\end{split}
\end{equation}
for $t_{nad}$. We then examine whether the frequency nadir constraint:
\begin{align}\label{con:sumfrequenynadir}
  \Delta \omega (t_{nad}(\mathcal{R}^{DERact}_{(k)},\mathcal{T}^{DERact}_{(k)},\mathcal{R}^{LDact}_{(k)},\mathcal{T}^{LDact}_{(k)})) \leq \Delta \omega^{max}
\end{align}
is satisfied in the current iteration. If constraint~\eqref{con:sumfrequenynadir} is satisfied, then the iteration process stops and the FFR reserve portfolio is obtained. Otherwise, the algorithm iteratively adds reserves from the lowest latency to the highest latency until constraint~\eqref{con:sumfrequenynadir} is satisfied. The overall procedure of heuristic reserve allocation is provided in Algorithm~\ref{alg:reserve_allocation}.

\begin{algorithm}[!t]
\SetAlgoLined
\KwIn{Set of DERs $\mathcal{D}$, set of controllable loads $\mathcal{C}$, maximum reserve capacity $R_d^{DERmax}, R_c^{LDmax}$, frequency deviation limit $\Delta \omega^{max}$, initial DER activation set $\mathcal{D}^{act} = \emptyset$, initial CL activation set $\mathcal{C}^{act} = \emptyset$. }
\KwData{Real-time latency measurements for DERs $\tau_d^{DER}$ for $d \in \mathcal{D}$ and CLs $\tau_c^{LD}$ for $c \in \mathcal{C}$.}

\textbf{Step 1: Latency-aware multipath routing optimization \ref{WP3IVA}}

\textbf{Step 2: Step response equivalent modeling of DERs \ref{WP3IVB}}

\textbf{Step 3: Sort step response models in an ascending order of latency \ref{WP3IVC}}

\textbf{Step 4: Iterative reserve allocation}

\For{$k = 1$ to $|\mathcal{D}|+|\mathcal{C}|$}{
  \eIf{$1 \leq id_{(k)} \leq |\mathcal{D}|$}
{   $ \mathcal{D}^{act}_{(k)} \cup id_{(k)}$

    $\mathcal{R}^{DERact}_{(k)}=\{R_{da}^{DERmax}\mid da \in \mathcal{D}^{act}_{(k)}\}$

    $\mathcal{T}^{DERact}_{(k)}=\{\tau_{da}^{DER}\mid da \in \mathcal{D}^{act}_{(k)}\}$
}{
    \If{$|\mathcal{D}| < id_{(k)} \leq |\mathcal{D}|+|\mathcal{C}|$}
    {   $\mathcal{C}^{act}_{(k)} \cup \left(id_{(k)}-|\mathcal{D}|\right)$
    
        $\mathcal{R}^{LDact}_{(k)}=\{R_{ca}^{LDmax}\mid ca \in \mathcal{C}^{act}_{(k)}\}$

        $\mathcal{T}^{LDact}_{(k)}=\{\tau_{ca}^{LD}\mid ca \in \mathcal{C}^{act}_{(k)}\}$
    }
}
  Compute nadir time $t_{nad}$ numerically satisfying: 
  \begin{equation}
  \begin{split}
      &\Delta \dot{\omega}_{sum}^{DER} (\mathcal{R}^{DERact}_{(k)},\mathcal{T}^{DERact}_{(k)}, t_{nad}) +\\ &\Delta \dot{\omega}_{sum}^{LD} (\mathcal{R}^{LDact}_{(k)},\mathcal{T}^{LDact}_{(k)}, t_{nad})-
    \dot{\omega}_L (t_{nad}) = 0 \nonumber
  \end{split}
\end{equation}
  \eIf{constraint~\eqref{con:sumfrequenynadir} satisfied}{
      $\mathcal{R}^{DERact}=\mathcal{R}^{DERact}_{(k)},\mathcal{R}^{LDact}=\mathcal{R}^{LDact}_{(k)}$ 
      \textbf{Break}\
    }{
      Increment the index to the next lowest latency: 
      $k = k +1$
    }
}
\KwOut{Allocated reserves portfolio $(\mathcal{R}^{DERact},\mathcal{R}^{LDact})$.}
\caption{Lowest-latency routing and heuristic reserve allocation algorithm}
\label{alg:reserve_allocation}
\end{algorithm}
\subsection{Computation acceleration techniques}\label{WP3IVE}
To satisfy the steady-state frequency security constraint~\eqref{con:steadystatecon}, the sum of reserves from both DERs and CLs needs to be greater than the contingency event $\Delta P_L$. Therefore, the sum of reserves in the initial activation set for CLs $\mathcal{R}^{DERact}_{ini}$ and DERs $\mathcal{R}^{LDact}_{ini}$ should be at least equal to the contingency event $\Delta P_L$, namely:
\begin{equation}\label{WP3con:steadystate}
    |\mathcal{R}^{DERact}_{ini}|+|\mathcal{R}^{LDact}_{ini}| \geq \Delta P_L
\end{equation}
This enables the iteration to commence from a non-empty activation set of DERs and CLs, providing adequate low-latency reserves to meet the steady-state constraint~\eqref{WP3con:steadystate}. 

According to~\eqref{eq:finalfreqdynamic}, when the latencies for DERs and CLs remain constant, there is a linear relationship between the frequency response and the reserve output. Simulations further reveal a monotonic relationship between the frequency nadir and the latencies. Therefore, it is possible to determine the range in which $t_{nad}$ lies before employing the heuristic algorithm. That is, the maximum nadir time $t_{nad}^{max}$ and the minimum nadir time $t_{nad}^{min}$ are derived as follows:
\begin{align}
    \begin{split}
    &\Delta \dot{\omega}_{sum}^{DER} (\mathcal{R}^{DERact}_{ini},\mathcal{T}^{DERact}_{ini}, t_{nad}^{max}) +\\ &\Delta \dot{\omega}_{sum}^{LD} (\mathcal{R}^{LDact}_{ini},\mathcal{T}^{LDact}_{ini}, t_{nad}^{max})-
    \dot{\omega}_L (t_{nad}^{max}) = 0\\ 
    &\Delta \dot{\omega}_{sum}^{DER} (\mathcal{R}^{DER},\mathcal{T}^{DER}, t_{nad}^{min}) +\\ &\Delta \dot{\omega}_{sum}^{LD} (\mathcal{R}^{LD},\mathcal{T}^{LD}, t_{nad}^{min})-
    \dot{\omega}_L (t_{nad}^{min}) = 0
\end{split}
\end{align}
where $\mathcal{R}^{DER}$ and $\mathcal{R}^{LD}$ represent the complete sets of reserves from DERs and CLs, along with their respective latency sets $\mathcal{T}^{DER}$ and $\mathcal{T}^{LD}$. Thus, the numerical solver can accelerate the computation by constraining the search area to the interval $t_{nad} \in [t_{nad}^{min},t_{nad}^{max}]$ to determine the nadir time $t_{nad}$ using~\eqref{con:sumtadcalculate}.



\section{Case Studies}\label{Casestudies}
\subsection{Experiment settings}
\begin{table}[!b]
\renewcommand{\arraystretch}{1.2}
\caption{Basic setting of system parameters}
\label{WP3table:para}
\centering
\begin{minipage}{\linewidth}
\begin{center}
    \begin{tabular}{ll}
        \toprule
        \textbf{Parameter} & \textbf{Value} \\
        \hline
        Load damping ratio \( D \) & 0.1 \\
        System inertia constant \( H \) & 3 s \\
        Governor time constant \( T_g \) & 0.3 s \\
        Turbine time constant \( T_c \) & 0.5 s \\
        Reheat time constant \( T_r \) & 12 s \\
        Reheat gain \( F_h \) & 0.15 \\
        Storage-type DER time constant \( T_d \) & 0.1 s \\
        System droop characteristic \( K \) & 0.5 \\
\bottomrule 
\end{tabular}
\end{center}
\end{minipage}
\end{table}

The case studies aim to verify the effectiveness of the heuristic routing and reserve allocation algorithm and provide insights into impact of latencies on the frequency response. The base power is set to 1000 MW and the power related values are presented in per-unit (pu) base. First, the transfer function of the synchronous generators $\sum G_g(s)$ used in the case studies is derived from the aggregated multi-machine frequency response model \cite{shi2018analytical}:
\begin{align}
    \sum G_g(s) = \frac{F_h T_r s+1}{ (T_r s+ 1)(T_g s+ 1)(T_c s+1)}
\end{align}
Parameters and their default settings are given in Table~\ref{WP3table:para}. The system has up to 50000 storage type DERs and 50000 controllable loads. The time constants of the storage-type DERs are set to \( T_{d} = 0.1 \) s. The maximum reserve output from storage-type DERs $R_{d}^{DERmax}$ is generated randomly between $[10, 15]\cdot10^{-6}$ pu, whereas the maximum reserve output from CLs $R_{c}^{LDmax}$ is between $[1, 5]\cdot10^{-6}$ pu. The remuneration rate for storage-type DERs and CLs is set as \( C_{rr} = 0.025 \times 10^6 \) \$/p.u. The studied power mismatch incidents $\Delta P_L$ are between $[0.01, 0.12]$ p.u. depending on the number of the considered flexible devices. 

The activation latency data are derived from the real-world deployment of the SCION networks in seven countries, namely US, Brazil, Germany, Netherlands, Switzerland, Singapore, and South Korea. Using an active probing scheme, the one-way activation latencies between the AS assumed to be hosting the system operator and other ASs hosting DERs and CLs in one country are recorded for one week with a sampling granularity of 0.5 seconds. Then the data from seven countries are aggregated to provide the basis for the empirical distribution of the average communication latencies. The activation latency for both DERs and CLs is then randomly sampled from the empirical distribution of average latencies in the range of $[0, \tau_{\max}]$, where $\tau_{\max}$ denotes the maximum latency value in the discretized cumulative latency distribution.

The simulations of the electrical system are implemented in MATLAB. The solver fmincon is used to benchmark the heuristic algorithm against a commercial solver. All numerical evaluations were performed on a workstation equipped with an Intel i9-14900F CPU @ 2.00 GHz and 32 GB of RAM. 
\subsection{Communication Latency}
\begin{figure}[!t]
    \centering
      \subfloat[\label{fig:WP3CDF}]{%
        \includegraphics[width=0.45\linewidth]{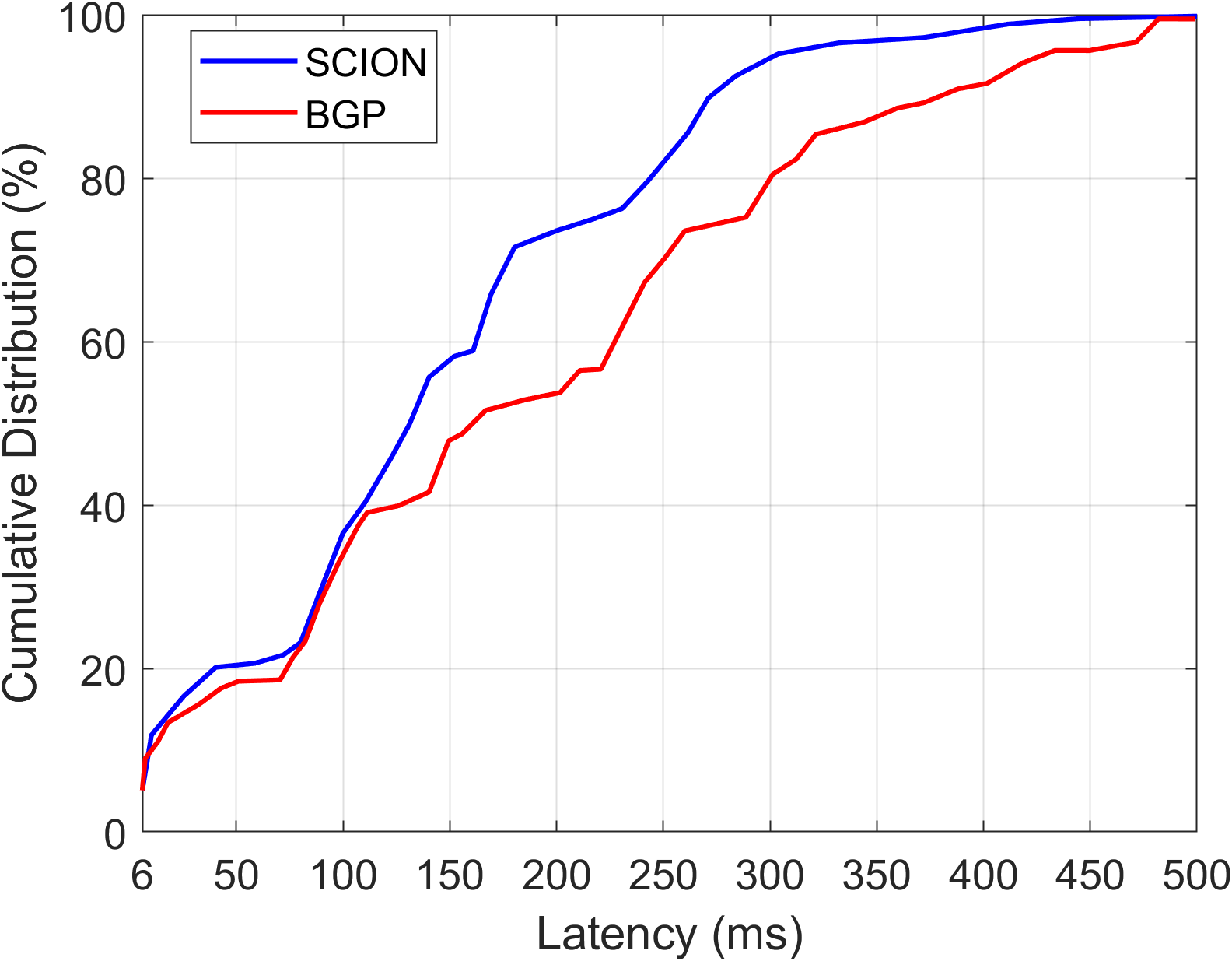}}
  \subfloat[\label{fig:WP3hist}]{%
       \includegraphics[width=0.45\linewidth]{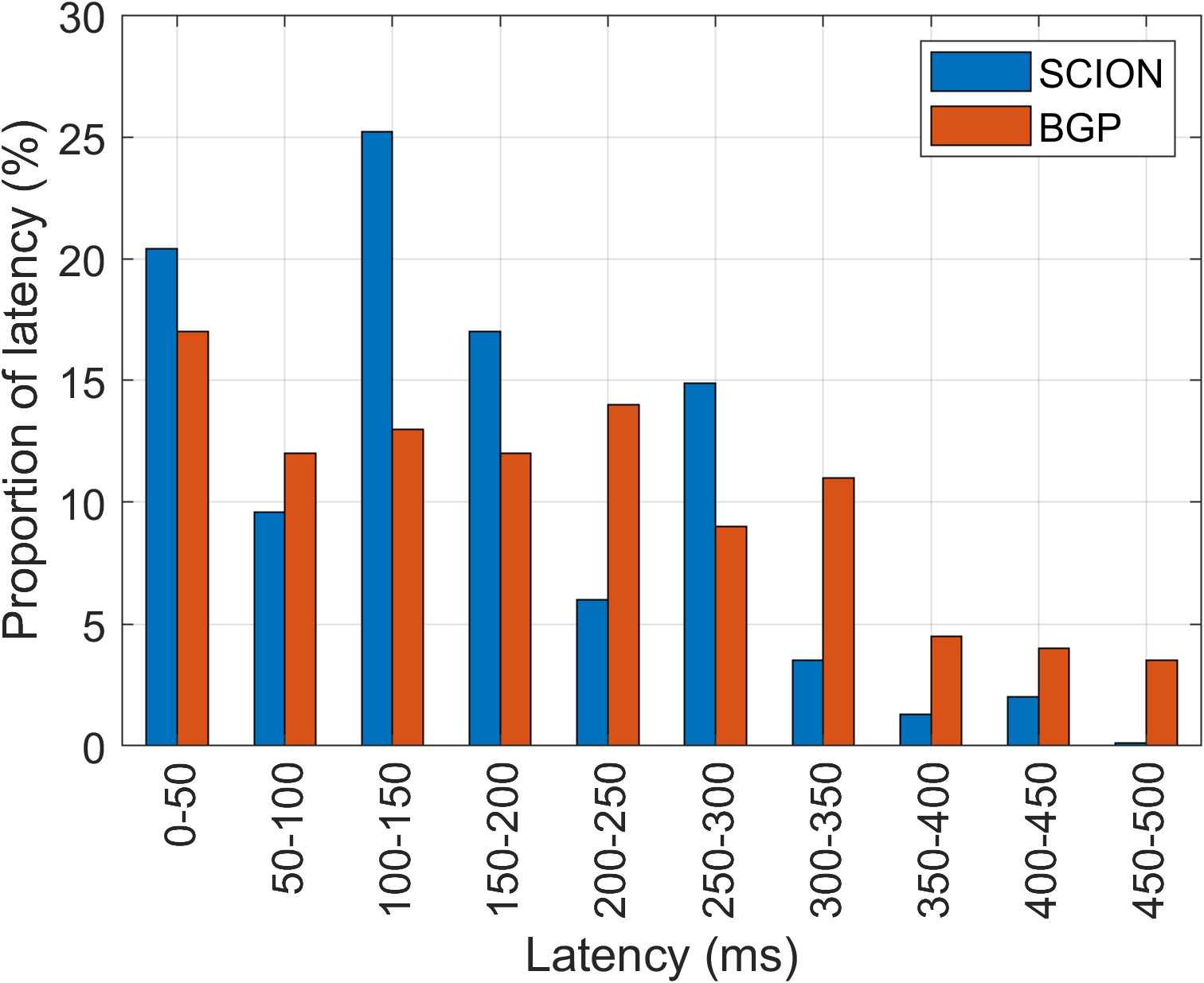}}
  \caption{\label{fig:HistSCIONBGPWP3} Comparison of the average communication latencies between the system operator and flexible loads with respect to the SCION-based Internet network and BGP-based Internet network measured by (a) empirical cumulative distribution function, and (b) histogram of latency measurements. }
\end{figure}
Figure~\ref{fig:WP3CDF} shows the empirical cumulative distribution functions (CDFs) of the one-way average communication latencies between the system operator and DERs and CLs simulated in seven countries using both SCION and BGP-based routing. The SCION network consistently achieves lower latencies in the upper tail of the distribution. Specifically, the 99th percentile latency under SCION is approximately 410 ms, while BGP exhibits a higher value of around 480 ms. This indicates that SCION reduces the highest communication delay by about 70 ms, which is significant for time-critical applications, such as FFR services. In addition, Figure~\ref{fig:WP3hist} reveals that SCION has a denser concentration of low-latency measurements (e.g., 100–200 ms bins), whereas BGP shows a more spread-out distribution, with more latencies lying in the 300–500 ms range. These results suggest that SCION-based Internet offers tighter bounds on the delay characteristics than BGP-based internet and enables a more responsive activation of DERs and CLs for FFR services.

\subsection{Optimality validation}
\begin{figure}[!t]
    \centering
      \subfloat[\label{fig:heuroptcostWP3}]{%
        \includegraphics[width=0.48\linewidth]{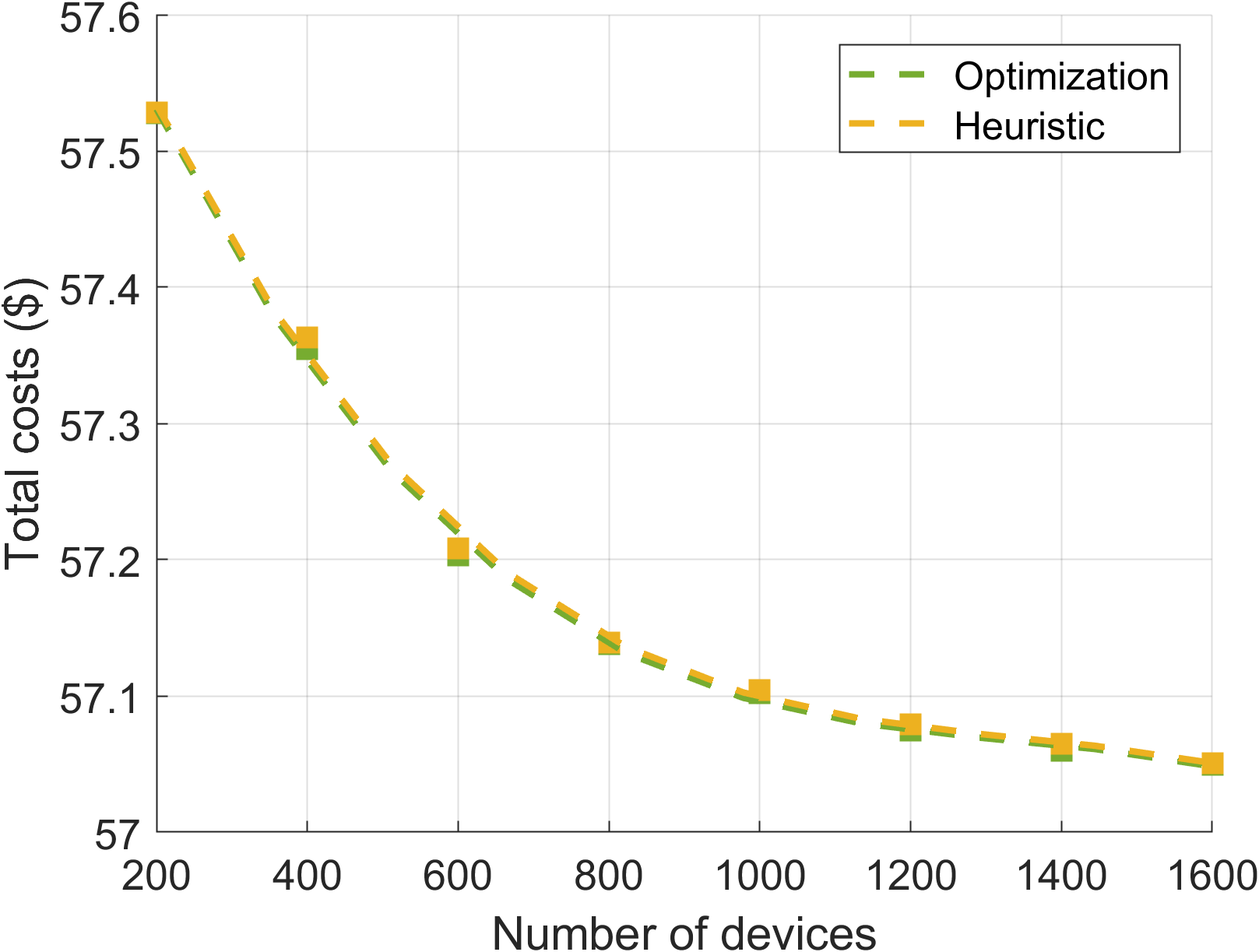}}
  \subfloat[\label{fig:heuroptfreqWP3}]{%
       \includegraphics[width=0.48\linewidth]{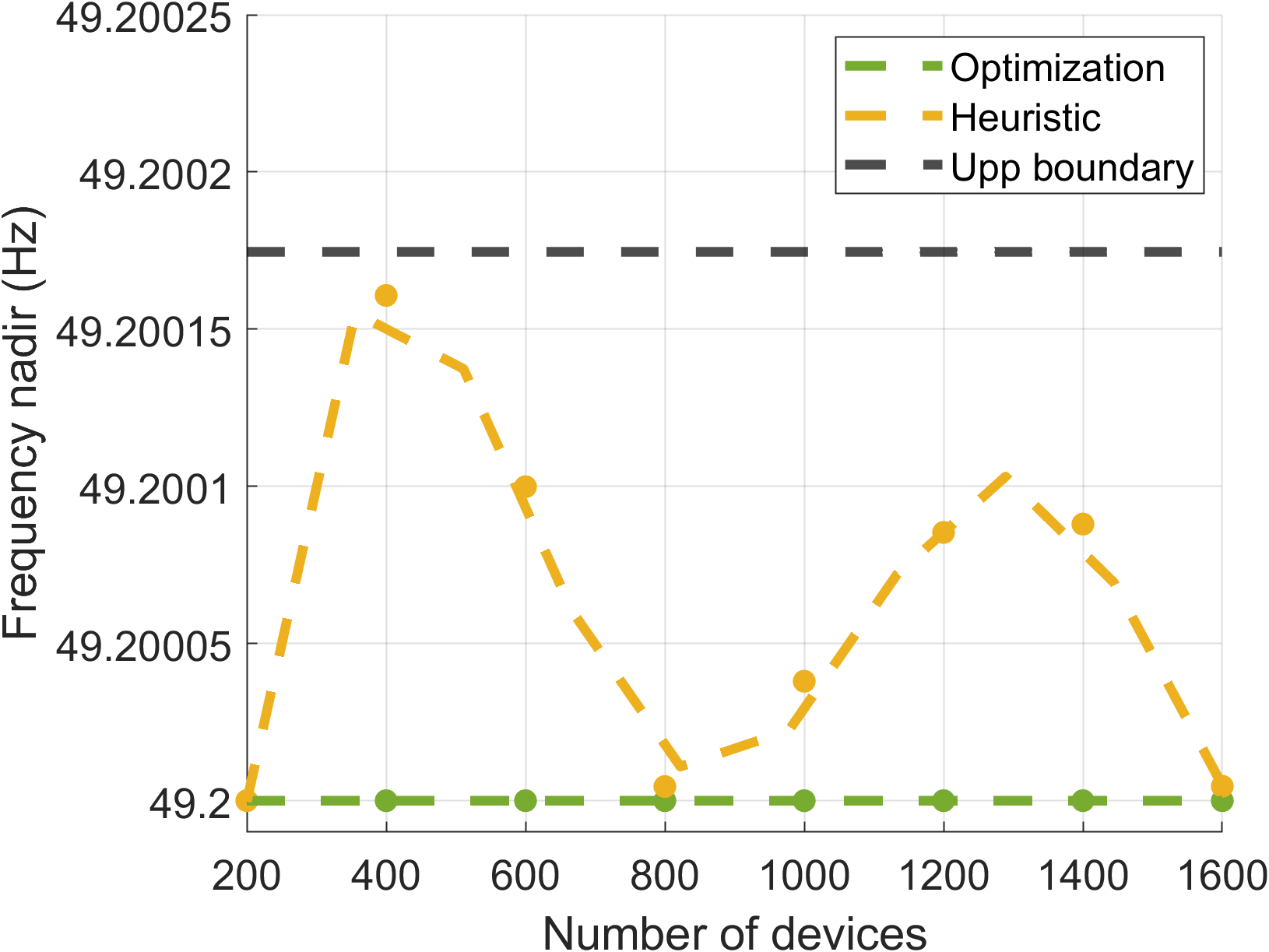}}
  \caption{\label{fig:heuropt} Comparison between the global optimization solver and the proposed heuristic algorithm evaluated by (a) the cost of FFR reserves and (b) the frequency nadir. }
\end{figure}


As a next step, we evaluate how accurately the heuristic algorithm performs in comparison to the standard nonlinear optimization solver for different numbers of flexible devices. Figure~\ref{fig:heuroptcostWP3} presents the total cost of the FFR service calculated by both the heuristic method and the optimization solver. The general trend indicates that the total cost for both methods decline as the available number of flexible devices increases. This occurs because as the number of available devices increases, the sampled data for communication latencies becomes more granular. When there is a limited number of available devices, there may be insufficient low latency devices to satisfy the frequency security constraint. As more low-latency reserves become available, both the heuristic algorithm and the optimization solver can satisfy the frequency security constraint with less total reserves, thereby lowering the total cost. The minor difference in the total cost between the heuristic algorithm and the optimization solver can be attributed to the discretization errors from the heuristic algorithm. As demonstrated in step 4 of Algorithm~\eqref{alg:reserve_allocation}, the reserve allocation method incorporates the maximum capacity from a flexible device into the FFR activation set during each iteration. In its final iteration, the heuristic algorithm includes the full capacity of a flexible device, leading to a surplus over the required reserve outputs and therefore resulting in a total cost slightly higher than the optimized result, as can be observed from Fig.~\ref{fig:heuroptcostWP3}. However, an important conclusion is that the expenses associated with FFR reserves identified by the heuristic algorithm align closely with those calculated using the global optimization solver.

Figure~\ref{fig:heuroptfreqWP3} shows the frequency nadir calculated based on the FFR output obtained from the heuristic model and the optimization solver. The optimization solver strictly ensures the frequency nadir to be at 49.2 Hz (the maximum frequency deviation limit), indicating that the frequency security constraint in problem~\eqref{WP3VPPoperation} is active. The heuristic algorithm, due to the discretization error between the maximum capacity activation and the optimized reserves, results in a slightly more conservative frequency nadir. It should be noted that the discretization error is bounded by the difference between the maximum capacity of the reserve in the last iteration and the optimal reserves of the flexible devices. In this case, the frequency response is determined by the maximum reserve capacity accessible from the set of all activated flexible devices. Using this information, we compute the upper boundary depicted by the black dashed line. Consequently, the heuristic algorithm is able to uphold frequency security even as the problem size increases by exhibiting a certain level of bounded conservativeness.

\subsection{Computation efficiency}
\begin{figure}[!t]
    \centering
      \subfloat[\label{fig:heuristicoptWP3}]{%
        \includegraphics[width=0.48\linewidth]{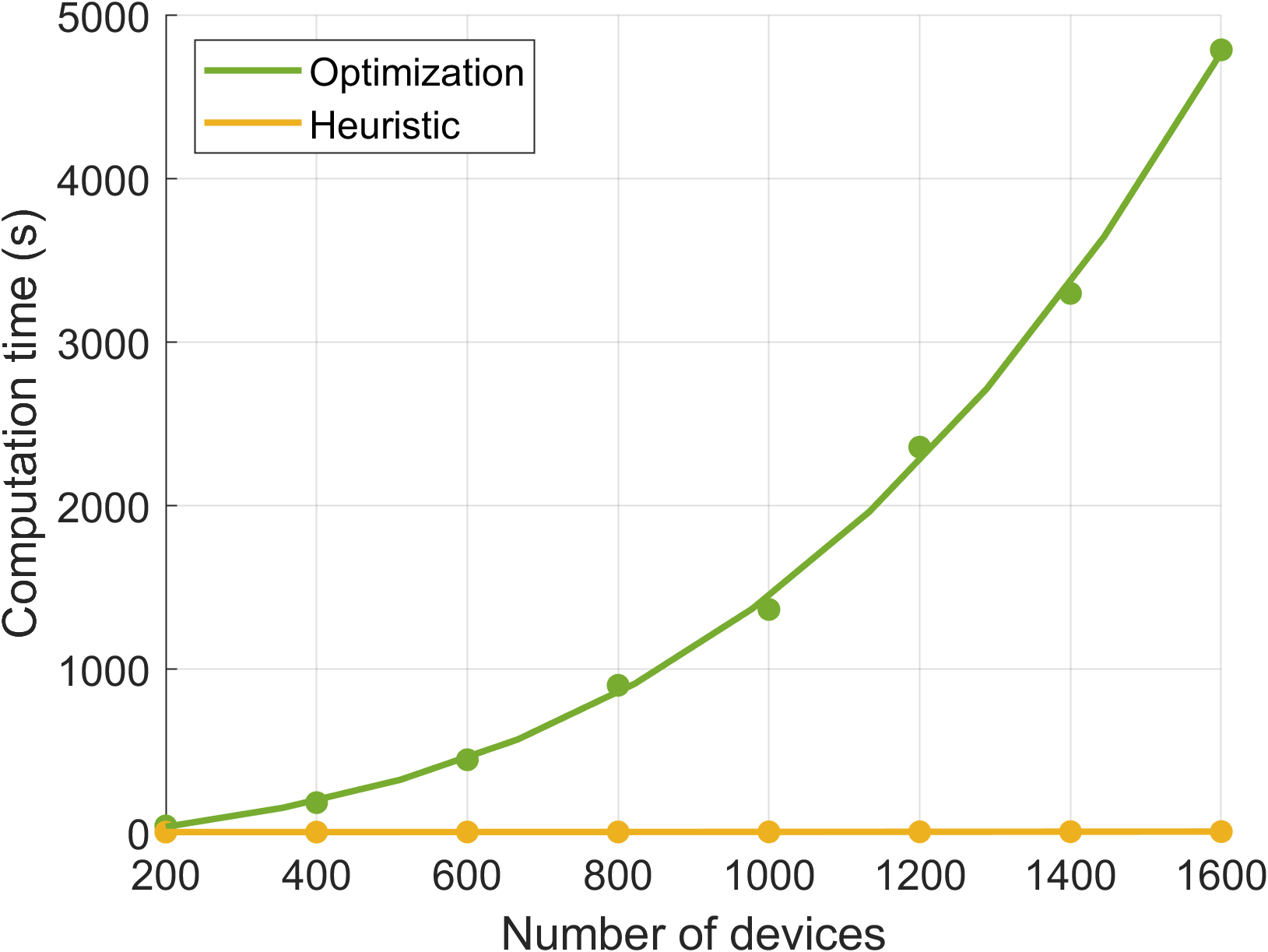}}
  \subfloat[\label{fig:heuristicWP3}]{%
       \includegraphics[width=0.48\linewidth]{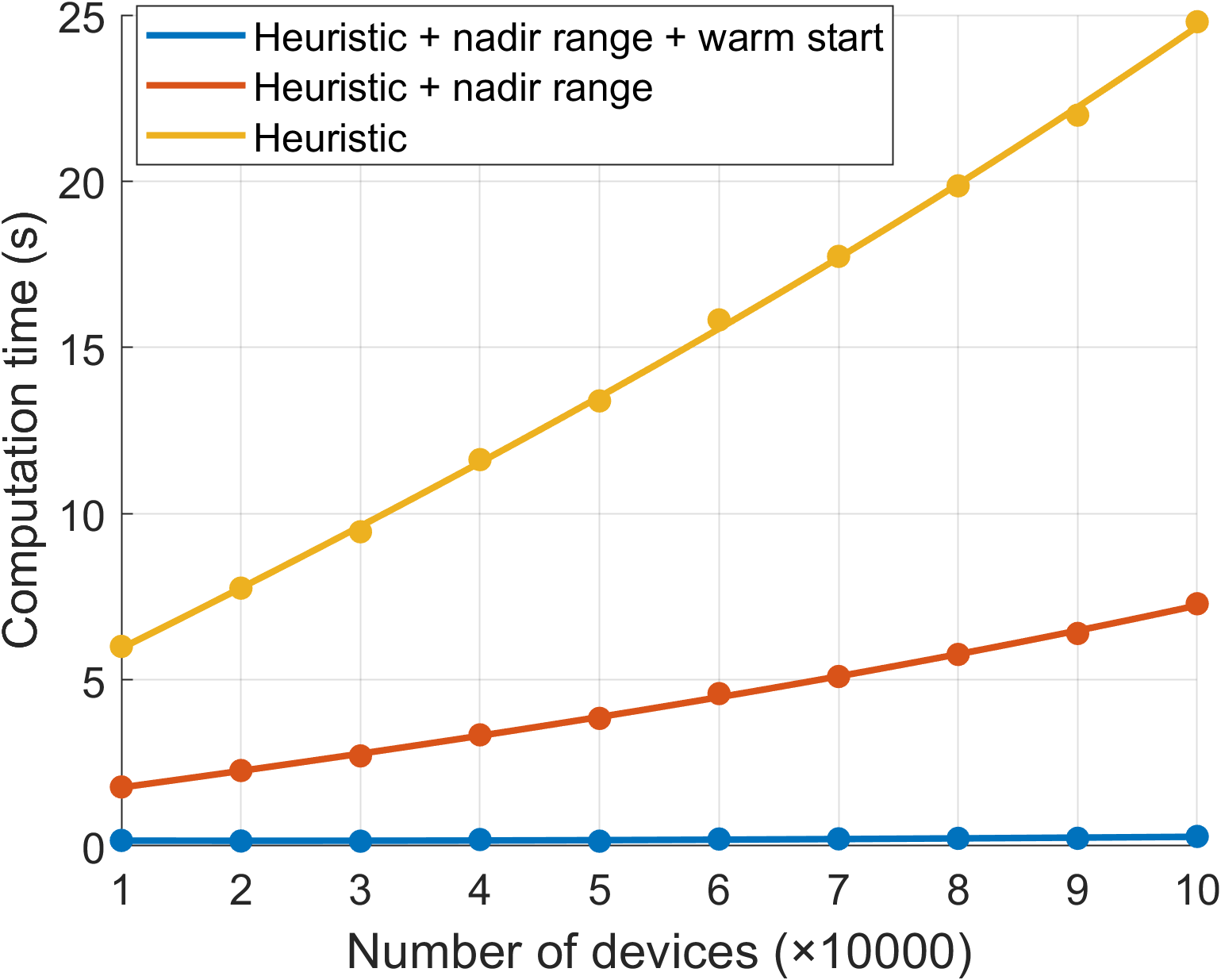}}
  \caption{\label{fig:heuroptcomputa} Computation efficiency compared between (a) the global optimization model versus the proposed heuristic model and (b) the heuristic model with computation acceleration techniques. }
\end{figure}

Figure~\ref{fig:heuristicoptWP3} shows the computation efficiency of the proposed heuristic algorithm compared to the fmincon optimization solver. The computation time required by the optimization solver rises exponentially. This is due to the NP-hard nature of the MINLP 
problem. As soon as the number of flexible devices exceeds 1600, the solver reports an error due to insufficient random access memory (RAM) of the PC. In constrast, the heuristic model exhibits a linear increase with the maximum latency in the test still being within seconds level. This is due to the heuristic algorithm's use of a directional and greedy search of reserve candidates, effectively avoiding the combinatorial complexity of the MINLP.   

With the implementation of additional search assistive approaches, the computation efficiency is further improved. As shown Fig.~\ref{fig:heuristicWP3}, the heuristic algorithm with nadir time estimation reduces the computation time by around 75\%. This is because the computation complexity of the heuristic algorithm lies mainly in the numerical solution of the frequency nadir time. As the range of the frequency nadir is narrowed, the search region for the root of the first derivative of the frequency nadir expression is accelerated. In addition, when the algorithm adopts warm start initialization based on the steady-state condition~\eqref{WP3con:steadystate}, the number of iterations for the algorithm is further reduced as the reserve outputs at the beginning of the search are closer to the optimized reserve outputs. As such, the computation time of the heuristic algorithm equipped with both nadir range prediction and the warm start technique is within 100 ms when the number of devices reaches up to 100000. This demonstrates the suitability and applicability of the proposed algorithm for real-time operation. 
\section{Conclusions}\label{conclusion}
In this paper, we present a SCION-based fast frequency response management strategy for the flexible devices including DERs and CLs in response to heterogeneous and time-varying communication delays. First, the time-domain frequency nadir expression is derived analytically incorporating the dynamic responses of both flexible devices and synchronous machines with droop control. Then, a cyber-physical co-optimization model is proposed which minimizes the operation cost of the FFR dispatch through both lowest latency routing and real-time reserve adjustments. To address the nonlinearity of the optimization model, a heuristic reserve allocation algorithm is proposed which approximates the optimal reserve portfolio thereby improving the computation efficiency. The results demonstrate the optimality and frequency security of the heuristic model. In addition, the computation efficiency is improved with the integration of both nadir prediction and computation acceleration techniques. Thus, the proposed strategy can be readily implemented by system operators for real-time adjustment of their reserve assets to counter dynamic communication latencies. Considering the time-critical aspect of FFR, the proposed SCION-based reserve dispatch architecture additionally allows a greater number of devices experiencing extended communication delays to become eligible for FFR services.

Future work can implement the proposed strategy using hardware-in-the-loop (HIL) simulation to evaluate its practicality and integration potential in real-world applications. In line with the cost-benefit principle, it is also recommended to conduct a minimum hardware requirement assessment to identify the most economical configuration for deploying the strategy on the real-world dispatch engines from system operators. 
\bibliographystyle{IEEEtran}
\bibliography{IEEEabrv, reference}

\section*{Appendix}
We first illustrate how to derive $\Delta \omega_L(t)$ based on the inverse Laplace transform of $\Delta \omega_L(s)$ from ~\eqref{eq:frequeninsep}. For simplicity, we define $a_u=\Re(\alpha_u),b_u=\Im(\alpha_u),c_u=\Re(p_u),d_u=\Im(p_u)$. For $\Delta \omega_L(s)$, we can distinguish three parts of sum:
\begin{align}\label{WP3appedmain}
\begin{split}   
  \Delta \omega_L(s)=&\frac{\Delta P_L}{s}\sum_{\substack{(\alpha_r, p_r)\in \mathcal{P}}} \frac{\alpha_r}{s+p_r} 
     \\&+\frac{\Delta P_L}{s}\sum_{\substack{(a_u+jb_u, c_u+jd_u)\in \mathcal{P}}}\frac{2a_u(s + c_u)}{(s + c_u)^2 + d_u^2}  \\&+\frac{\Delta P_L}{s}\sum_{\substack{(a_u+jb_u, c_u+jd_u)\in \mathcal{P}}}\frac{2b_ud_u}{(s+c_u)^2 + d_u^2}
\end{split}
\end{align}

To transform $\omega_L(s)$ into the time domain, we apply the inverse Laplace transform for each part individually:
\begin{align}
\begin{split}
   & \mathcal{L}^{-1}\left\{\frac{\Delta P_L}{s}\frac{\alpha_r}{s + p_r}\right\} = \Delta P_L\int_{0}^{t} \mathcal{L}^{-1}\left\{\frac{\alpha_r}{s + p_r}\right\} \,d\tau 
    \\
    &= \Delta P_L\alpha_r\int_{0}^{t} e^{-p_r\tau} \,d\tau
    \\
    &= \Delta P_L\frac{\alpha_r}{p_r}\left( 1 - e^{-p_rt} \right) 
\label{eq:w_lexp}
\end{split}
\end{align}

\begin{align}
\begin{split}
    & \mathcal{L}^{-1}\left\{\frac{\Delta P_L}{s}\frac{2a_u(s+c_u)}{(s + c_u)^2 + d_u^2}\right\}= \Delta P_L\int_{0}^{t} \mathcal{L}^{-1}\left\{\frac{2a_u(s+c_u)}{(s + c_u)^2 + d_u^2}\right\} \,d\tau ,
    \\
    &= \Delta P_L 2a_u\int_{0}^{t} e^{-c_u\tau}cos(d_u\tau) \,d\tau
    \\
    \\
    &= \Delta P_L\frac{2a_u}{c_u^2 + d_u^2}\Big[-e^{-c_ut}cos(d_ut)c_u + c_u + e^{-c_ut}sin(d_ut)d_u\Big]
\label{eq:w_lcos}
\end{split}
\end{align}

\begin{align}
\begin{split}
    &\mathcal{L}^{-1}\left\{\frac{\Delta P_L}{s}\frac{2b_ud_u}{(s+c_u)^2 + d_u^2}\right\} = \Delta P_L\int_{0}^{t} \mathcal{L}^{-1}\left\{\frac{2b_ud_u}{(s+c_u)^2 + d_u^2}\right\} \,d\tau ,
    \\
    &= \Delta P_L2b_u\int_{0}^{t} e^{-c_u\tau}sin(d_u\tau) \,d\tau
    \\
    \\
    &= \Delta P_L\frac{2b_u}{c_u^2 + d_u^2}\Big[-e^{-c_ut}sin(d_ut)c_u + d_u- e^{-c_ut}cos(d_ut)d_u\Big]
\label{eq:w_lsin}
\end{split}
\end{align}

Substituting \eqref{eq:w_lexp} to \eqref{eq:w_lsin} into~\eqref{WP3appedmain}, we obtain the expression for $\Delta \omega_L(t)$ in~\eqref{eq:finalfreqdynamic}. 

Since \( \Delta \omega_c^{LD}(s) \) in~\eqref{eq:frequeninsep} is simply \( \Delta \omega_L(s) \) multiplied by a CL's reserve capacity $R_c^{LD}$ and its time-delay exponential term $e^{-s \tau_c^{LD}}$, we can apply the Laplace transform property for time-delayed functions:
\begin{equation}
    \mathcal{L}\left\{ u(t - t_0) f(t - t_0) \right\} = e^{-t_0 s} F(s)
    \label{eq:time_delay}
\end{equation}
Using this property, the derivation of \( \Delta \omega_c^{LD}(t) \) follows a similar pattern as the steps in~\eqref{eq:w_lexp} to~\eqref{eq:w_lsin}, with the only differences being the time shift \( \tau_c^{LD} \) and the scaling factor \( R_c^{LD} \).

The final step is to transform \( \Delta \omega_d^{DER}(s) \) into the time domain. To facilitate an easier inverse Laplace transformation, the additional pole at \( -\frac{1}{T_d} \) obtained from the term $1/(sT_d+1)$ is incorporated into the residue decomposition process given in~\eqref{eq:WP3normalpoles}. On this basis, we define a new set of poles and corresponding residues denoted as \( (\alpha_i, p_i) \in \mathcal{P}^{DER} \):
\begin{align}\label{eq:sumfirstorderT_d}
\begin{split}
    \frac{\Delta \omega(s)}{\Delta P(s)(sT_d + 1)} &= \sum_{(\alpha_i, p_i)\in \mathcal{P}^{DER}} \frac{\alpha_i}{s + p_i}
    \\
     \sum_{d\in \mathcal{D}} \Delta \omega_d^{DER}(s)& = \sum_{d\in \mathcal{D}}\sum_{(\alpha_i, p_i)\in \mathcal{P}^{DER}} \frac{R_d^{DER}e^{-s \tau_d^{DER}}}{s} \cdot \frac{\alpha_i}{s + p_i}
\end{split}
\end{align}

As a result, the derivation of \( \Delta \omega_L(t) \) can be adapted for \( \Delta \omega_d^{\mathrm{DER}}(t) \) by incorporating the time delay \( \tau_d^{\mathrm{DER}} \) and the reserve capacity \( R_d^{\mathrm{DER}} \) of DER \( d \).

\ifCLASSOPTIONcaptionsoff
\newpage
\fi
\end{document}